# Giant g factor tuning of long-lived electron spins in Ge

Anna Giorgioni[1]*, Stefano Paleari[2], Stefano Cecchi[3†], Emanuele Grilli[1], Giovanni Isella[3], Wolfgang Jantsch[4], Marco Fanciulli[2], Fabio Pezzoli[1]*

1– LNESS and Dipartimento di Scienza dei Materiali, Università di Milano Bicocca, via Cozzi 55, 20125 Milano, Italy
2 – Dipartimento di Scienza dei Materiali, Università di Milano Bicocca, via Cozzi 55, 20125 Milano, Italy
3 – LNESS and Dipartimento di Fisica, Politecnico di Milano, via Anzani 42, 22100 Como, Italy
4 – Institut für Halbleiter- und Festkörperphysik, Johannes Kepler University, Altenbergerstrasse 69, 4040 Linz, Austria
[†] current address: Paul-Drude-Institut für Festkörperelektronik, Hausvogteiplatz 5-7, 10117 Berlin, Germany
* anna.giorgioni@unimib.it; fabio.pezzoli@unimib.it

**Control of electron spin coherence via external fields is fundamental in spintronics. Its implementation demands a host material that accommodates the highly desirable but contrasting requirements of spin robustness to relaxation mechanisms and sizeable coupling between spin and orbital motion of charge carriers. Here we focus on Ge, which, by matching those criteria, is rapidly emerging as a prominent candidate for shuttling spin quantum bits in the mature framework of Si electronics. So far, however, the intrinsic spin-dependent phenomena of free electrons in conventional Ge/Si heterojunctions have proved to be elusive because of epitaxy constraints and an unfavourable band alignment. We overcome such fundamental limitations by investigating a two dimensional electron gas (2DEG) confined in quantum wells of pure Ge grown on SiGe-buffered Si substrates. These epitaxial systems demonstrate exceptionally long spin relaxation and coherence times, eventually unveiling the potential of Ge in bridging the gap between spintronic concepts and semiconductor device physics. In particular, by tuning spin-orbit interaction via quantum confinement we demonstrate that the electron Landé *g* factor and its anisotropy can be engineered in our scalable and CMOS-compatible architectures over a range previously inaccessible for Si spintronics.**

Spin-orbit interaction (SOI) couples charge and spin degrees of freedom [1]. This effect has sparked considerable interest because it results in a suitable spin splitting even in the absence of external magnetic fields. SOI governs spin-dependent phenomena such as Bychkov-Rashba physics [2,3,4], persistent spin helix states [5,6,7], spin Hall [8,9,10] and spin Seebeck effects [11,12], offering novel and exciting perspectives for utilizing spin currents in non-magnetic materials [13]. This opens viable avenues for the end-of-the-roadmap implementation of semiconductor spintronics [14].

The Landé $g$ factor describes the susceptibility of the spin state of a charge carrier to an external field and sets a key metrics for the strength of SOI in the solid state framework. In Si, the fundamental building block of mainstream microelectronics, the weak SOI manifests itself as a negligible deviation of the electron $g$ factor from the isotropic free carrier value $g_0 \approx 2$ [15]. Spin-orbit coupling due to bulk inversion asymmetry is in fact absent in Si due to its centrosymmetric crystal structure [1,16]. Seminal works demonstrating tailoring of SOI in Si rather focused on low-dimensional Si/SiGe heterosystems, in which SOI becomes more important at the interfaces as a result of the induced spatial inversion asymmetry [17,18,19]. Yet the $g$ factor tunability in such systems remained very small [20].

In this contest, we turned our attention to Ge because it exhibits a highly anisotropic $g$ factor [21] due to a stronger SOI than the lighter Si, while it shares with Si the key prerequisites for any practical implementation of quantum information processing, namely a long spin relaxation time and a substantial abundance of spin-less isotopes [22,23]. In view of its full compatibility with the technology of integrated circuits and its exceptionally high bulk mobility, Ge is also increasingly seen as a viable option for replacing Si in conventional low-power logics [24] and can thus be regarded as an attractive candidate for transport in novel spintronic architectures.

Recently, manifold and intriguing phenomena have been revealed in Ge-based heterostructures. Cubic-k terms have been shown to dominate the $k \cdot p$ SOI Hamiltonian of two-dimensional hole gases [25]. Electric-field-induced tuning of the hole $g$ factor [26] has been reported in hybrid devices made from superconductors and self-assembled nanocrystals [27], while core-shell Ge/Si nanowires [28] have been envisioned as hosts of Majorana fermions [29].

To date, however, efforts have been mainly focused on the spin physics of holes. Besides the large lattice mismatch, which induces growth defects and poor material and interface quality, the spontaneous type II band alignment at Ge/Si heterojunctions [30,31] has so far precluded the experimental study of SOI of conduction electrons confined in Ge. Indeed charge carriers are spatially separated by the built-in potential, which favours holes (electrons) at the Ge (Si) side of the heterointerface.

In light of the pivotal advances reported in the field of Si photonics [32,33], we expect that band-gap engineering in SiGe alloys will similarly provide tremendous advantages to semiconductor spintronics by opening unexplored pathways for the full exploitation of Ge. The vast degrees of freedom offered by strain and alloying in dictating the band-edge offsets in SiGe heterostructures motivated us to design n-type modulation (n-mod) doped devices on Si that consist of pure Ge quantum wells (QW) embedded in Ge-rich SiGe barriers with a 10 nm thick phosphorous doped region at their centre (Fig.1a). The individual layers were engineered in order to obtain negligible strain with respect to the SiGe buffer, as confirmed by high resolution x-ray diffraction (HRXRD) measurements summarized in Fig.1b,c and in the Supplementary Section A. Such strain-compensation accommodates the compressed QW within tensely strained barriers and precludes the formation of additional defects at the interfaces. The resulting Ge/SiGe heterojunction allows us to gather direct access to a type I band alignment, with a notable accumulation of *L*-valley electrons (see Fig.1d and Fig.2a) in the Ge well due to a robust confining potential of the order of 60 meV. This, combined with conduction electron spin resonance (CESR), permits experimental manipulation of the electron $g$ factor theoretically predicted in Ge more than a decade ago [34]. To this end, we carried out a systematic study of samples that, according to HRXRD, differ by the QW thickness, namely 20±1, 17±1 and 16±1 nm.

We found that the low temperature cyclotron resonance (CR) strongly depends on the relative orientation of an external magnetic field $\boldsymbol{B}$ with respect to the sample surface. As shown in Fig.2b, the sample with the largest width of the QWs and without remote doping does not show a CR signal when $\boldsymbol{B}$ lies along the [110] direction (in-plane field). On the other hand, when $\boldsymbol{B}$ is rotated towards the [001] growth direction (perpendicular field), the spectrum exhibits a very pronounced signal. The resonance field decreases as $\theta \to 0$ according to $1/\cos\theta$ (see Supplementary Section B) [17]. This behaviour is a clear signature that carriers are confined in the (001) plane, where they can undergo cyclotron motion driven by the electric field of the microwave. This well-defined CR and its characteristic dependence upon illumination, discussed in detail in the Supplementary Section B, provides direct proof of the existence of a 2DEG in the QWs plane [17], and the absence of low temperature localization of carriers on impurity sites.

As a consequence, we have direct access to the intrinsic spin-dependent properties of conduction electrons. This constitutes a remarkable difference with respect to previous electron spin resonance studies applied to Ge. Aside works focussed on electrons bound to donors, very few experiments suggested the peculiar presence of an ESR due to delocalized electrons in antimony-doped bulk Ge at low temperatures [35,36,37]. Such finding was ascribed to partial population of conduction band states by the built-in inhomogeneous strain fields randomly experienced by electrons at different Sb sites. Instead, our heteroepitaxial n-mod architecture naturally guarantees itinerant electrons in the Ge layer and their concomitant spatial separation from the remote donors, that reside in the SiGe barrier.

This point is further corroborated by the following results. In addition to the CR signal, Fig.2c shows that four well-resolved CESR peaks become prominent in n-mod samples. These peaks (A to D in Fig.2c) strongly shift in their spectral position when increasing the angle $\theta$ between $\boldsymbol{B}$ and the normal to the sample surface from 0° to 90°. This dependence demonstrates a highly anisotropic $g$ factor, as summarized in Fig.2d. Notably, we did not succeed in observing peaks corresponding to electrons localized on P donors, neither in the SiGe barriers nor in the Ge wells. The origin of the narrow resonance lines shown in Fig.2c and of their marked angular dispersion can be rationalized as detailed below.

In bulk Ge each conduction band edge at the four equivalent $L$ points of the Brillouin zone has an ellipsoidal energy surface oriented along a <111> crystal direction (see Fig.2a). According to Roth and Lax [21], the $g$ factor matrix of free electrons reflects such spheroidal shape and its $C_{3v}$ symmetry [35]. Hence for any angle θ between the external field and the major axis of one ellipsoid of revolution, the concomitant effective value of $g$ can be obtained as follows:

$$g^2 = g_p^2 \cos^2 \phi + g_t^2 \sin^2 \phi \qquad (1)$$

where $g_p$ and $g_t$ are the two independent parallel and transverse components lying along or being normal to the major axis of the ellipsoid, respectively. Here $\phi$ is the angle between **B** and the major axis of the ellipsoid [38].

Fig.2d compares our experimental data for QWs and the angular dependence of the $g$ factor of conduction electrons in bulk Ge as obtained by using equation (1) and the $g_p$ and $g_t$ values from Refs. 36,39 (dotted lines). Their striking agreement demonstrates, at a glance, that the CESR features of Ge wells originate from itinerant $L$-valley electrons and that the heterostructures preserve the bulk $C_{3v}$ symmetry of the $g$-tensor. Such finding is in sharp contrast to the behaviour of the magneto-conductivity tensor, which rules the CR response (Fig.2b). This might be a consequence of the fact that the latter is mostly determined by heterointerface properties, while the $g$ factor deviations from the free electron value are caused first of all by SOI [17]. A theoretical approach will be mentioned below. It is worth noting that the observation of a well-resolved CESR multiplet proves that spin relaxation of conduction electrons in QWs is dominated by zone-centre *intravalley* rather than zone-edge *intervalley* electron-phonon coupling [40]. The latter due to scattering among the different $L$ minima would have otherwise averaged out the $g$ factors, eventually yielding a single CESR line [41]. We emphasize that the inversion symmetry of the Ge lattice is well-known to preclude D'yakonov-Perel type spin-flip processes so that spin relaxation is essentially mediated by Elliott-Yafet mechanisms. This feature and the unique SOI experienced by thermal electrons at the conduction band edge of Ge have been recently suggested resulting in exceptionally long-lived electron spin states [40,42]. By working at cryogenic temperatures, we could selectively quench the intervalley scattering, that previous literature work recognized as a major factor in limiting the experimentally accessible spin relaxation times [39,43]. Such approach opens up a largely unexplored scenario offering the possibility to capture ultimate spin-flip and dephasing mechanisms.

Fig.2d allows us to obtain via equation (1) the $g_p$ and $g_t$ values for the QWs (solid lines in) and to identify the pristine valleys (Fig.2a) giving rise to the observed resonance lines. To better appreciate this, we can start considering θ = 0° where all the CESR peaks merge at $g$∼1.66. The weak removal of degeneracy, which can be noticed in the QW data of Fig.2d, is due to a small but unavoidable misorientation of the sample of ∼0.5° towards the [110] direction. By increasing θ to ∼55°, the external field aligns with the major axis of the ellipsoidal energy surface of valley B, highlighted in red in Fig.2a, and its associated $g$ factor decreases to the minimum value, that is $g_p$. Instead, for valleys C and D (blue in Fig.2a) a 90° increase of θ yields an increase of $g$ from 1.66 up to the largest value, namely $g_t$.

A closer look to Fig.2d already points out that at a fixed θ, the $g$ factor of bulk and QWs might differ. In particular, the mismatch is maximum when $g = g_p$ and vanishes when $g = g_t$. Those changes can indeed be used as sensitive probes of the electronic band structure being the intertwined surge of strain and confinement effects on SOI [34]. In this work we were able to disentangle these two contributions by focussing entirely on the latter. In fact, while shaping confinement of the electron wavefunctions via the QW width, all our heterostructures retain the same strain level being set by the lattice mismatch between Ge and the buried SiGe buffer (see also data in Supplementary Section A).

Fig.3a reports $g_p$ and $g_t$ as a function of the well thickness (diamonds) along with the corresponding bulk Ge benchmarks (arrows) taken from the literature [36]. Remarkably, while $g_t$ of bulk and QW coincide within the experimental error, $g_p$ becomes substantially larger than the bulk limit as the QW width decreases. The findings summarized in Fig.3 constitute the experimental proof of a puzzling SOI effect induced by interactions between the lowest conduction band at the *L* point and the other close and remote bands. Such phenomena were unveiled by $k \cdot p$ perturbation theory by Baron and coworkers [34], who anticipated the renormalization of the *g* factor of *L*-valley electrons in Ge/SiGe QWs. The excellent agreement between theory and experiments can also be noticed in Fig.3b, where our data at θ = 90° for degenerate A and B valleys (dots) are directly compared with the corresponding calculations from Ref. 34 (solid line).

We emphasize that, although electron *g* factor manipulation has been largely addressed in QWs of III-V compounds [44,45], our approach discloses a giant shift directly in group IV materials. Here we leverage SOI shaping in low dimensional structures to extend the *g* factor tunability by more than one order of magnitude compared to the experimental works on Si-based systems published to date [20,46,47].

CESR investigation of the anisotropic electron dispersion of *L*-valleys provides crucial insights also into the electron spin coherence. To address this further, we now focus on the CESR lineshape in an attempt to identify the homogeneous Lorentzian linewidth $\Delta B^0{}_{pp}$ and possible broadening mechanisms of the resonance peaks that might conceal transverse spin relaxation processes [48]. Fig.4a reports CESR lines corresponding to various *g* factors measured in the sample with the widest QWs. For a better comparison, the spectra are shifted by an amount equal to their own resonance field. Unexpectedly, the measured peak-to-peak linewidth $\Delta B_{pp}$ decreases when the *g* factor increases. These peaks belong to either valley A or B. However, the upper panel of Fig.4b summarizes similar results also for valleys C and D, thus clarifying that $\Delta B_{pp}$ does not depend upon the valley index, but it is exclusively linked to the value of *g*. The observation of a similar broadening for peaks originating from independent valleys provides further evidence of the pivotal role played by the intravalley relaxation.

To gather better insight about decoherence mechanisms, we start noticing that the epitaxial growth of strained Ge layers is always accompanied by surface roughness, yielding fluctuations of the QW width. On the time scale of momentum relaxation within a valley, the electron spin can experience scattering through regions of randomly changing $g$ factor, which reflect the in-plane variations of the well thickness. As a consequence, the spin state of the electron dephases providing a relevant source of inhomogeneous Gaussian broadening $\Delta B^G{}_{pp}$ of the CESR peaks. As detailed in the Supplementary Section C, we evaluated this contribution by using the root-mean-square roughness of the sample surface as obtained by Atomic Force Microscopy (inset of Fig.4b). The latter is ~2nm for all the QW samples. The results of this analysis are displayed as an orange line in the upper panel of Fig.4b, and highlight two interesting points. First, $\Delta B^G{}_{pp}$ vanishes at high $g$ values. In particular, $\Delta B^G{}_{pp} \equiv 0$ when $g = g_t$, which explains why the measured CESR peaks at $g_t$ (see an example in Fig.4a) are the only ones to be well approximated by a Lorentzian curve (red line in Fig.4a). Second, $\Delta B^G{}_{pp}$ gains weight when $g$ decreases. However, the not complete agreement with the experiments evidences that, although interface roughness is the main source of CESR broadening, other contributions have to be taken into account to fully explain the observed lineshapes. In particular, we notice that to a first approximation the experimental data can be recovered by applying a rigid shift of ~2.9 G to the calculated $\Delta B^G{}_{pp}$ (compare solid and dotted lines in Fig.4b). Since this offset points towards isotropic decoherence mechanisms, we suggest the following scenario to explain the physics leading to such 2.9 G broadening.

The stochastic nature of the diffusion process washes out the inhomogeneous magnetic fields arising from the nuclear spins of the naturally occurring $^{73}$Ge isotopes. Hence itinerant $L$-valley electrons experience an effective suppression of the hyperfine relaxation and are expected to yield the so-called motional narrowing, i.e. a reduced CESR linewidth. Nevertheless, during their random walk within the QW plane, mobile electrons are likely to reside for a finite time in smooth potential islands induced by thickness fluctuations prior to jump into a neighbouring in-plane site. This weak localization enhances the Fermi contact interaction between the spin and the local nuclear fields, sustaining dephasing and, in turn, CESR broadening. The isotropic component of the linewidths pointed out in Fig.4b can thus be accounted for by the two aforementioned opposing effects, namely hyperfine coupling and motional narrowing. In our Ge QWs the prominent role of the latter leads to a remnant hyperfine broadening of 2.9 G. This is substantially narrower than the 10 G resonance linewidth of electrons fully bound to shallow donors that is well known for bulk samples with natural isotopic abundance of $^{73}$Ge [49,22].

It shall be noted that the phenomenon discussed above neglects broadening due to spin-flip processes, consistently with the spin relaxation times addressed in the following.

Data in the upper panel of Fig.4b further show the occurrence of slightly different linewidths at the same $g$ value, thus suggesting the presence of additional, albeit weaker, dephasing mechanisms. With this respect, it is illuminating to note that we measured two $g_t$ peaks: One at $\theta \sim 35°$ (Fig.2d), having $\Delta B_{pp} = 5.8 \pm 0.4$ G, and the other at $\theta = 90°$, having $\Delta B_{pp} = 2.9 \pm 0.1$ G. Similar $\Delta B_{pp}$ ratios of these two CESR lines have been systematically observed in all the QW samples, thus highlighting that the transverse spin relaxation is systematically slower when the external magnetic field is along the QW plane, i.e. $\theta = 90°$. Such finding can be understood by considering dephasing mechanisms that arise within the dominant Elliott-Yafet framework [40] (see Supplementary Section D). The Elliott-Yafet spin relaxation can be described by a random effective magnetic field $\boldsymbol{B_{EY}}$ acting on the electron spin at each scattering event. In intravalley processes, which we showed to be the main ones in the analysed system, $\boldsymbol{B_{EY}}$ lies along the direction defined by the cross product $\boldsymbol{k} \times \boldsymbol{k'}$, where $\boldsymbol{k}$ and $\boldsymbol{k'}$ are the momenta of the initial and final electron states in the scattering event. In a 2DEG this results in $\boldsymbol{B_{EY}}$ along the growth direction, thus providing transverse relaxation of in-plane spin components, namely the one probed at $\theta = 0°$, but it does not affect out-of-plane spin components, i.e. θ = 90° configuration. As a consequence, as θ decreases towards 0°, the Elliott-Yafet relaxation becomes more important, manifesting itself in our experimental data as a sizeable contribution to the broadening of CESR lines.

The observation of larger linewidths at small θ values also rules out decoherence due to Bychkov-Rashba SOI. Although precluded by the symmetric design of our n-mod structures, this effect can still possibly occur because of the rotoinversion asymmetry induced by the finite, unavoidable roughness of the interfaces [50]. The nature of such spin-orbit coupling, if any, would lead to a Rashba field oriented within the 2DEG plane [3] and would provide an additional channel of transverse spin relaxation that, as opposed to our findings, increases the linewidth when θ approaches 90° [51].

After having discussed all the mechanisms contributing to the observed CESR linewidth, we can determine the relaxation time of the spin ensemble $T_2^*$, which provides a lower limit for the spin decoherence time $T_2$ [48], as follows:

$$T_2^* = \frac{\hbar}{g\mu_B} \frac{2}{\sqrt{3}\, \Delta B^0{}_{pp}} \qquad (2)$$

where ℏ is the reduced Planck constant, $\mu_B$ the Bohr magneton, $g$ is obtained from the CESR peak, and $\Delta B^0{}_{pp}$ can be obtained by the following relation [52]:

$$\Delta B_{pp}(g) = \frac{1}{2} \Delta B^0{}_{pp} + \sqrt{\frac{1}{4}\, [\Delta B^0{}_{pp}]^2 + [\Delta B^G{}_{pp}(g)]^2} \qquad (3)$$

using the measured $\Delta B_{pp}(g)$ shown in Fig.4b, and the inhomogeneous broadenings $\Delta B^G{}_{pp}(g)$ as calculated in the Supplementary Section C.

The values of $T_2^*$ for the widest QW sample are summarized in the lower panel of Fig.4b. Similar data have been found also for narrower QWs (Supplementary Section C). In agreement with the physical picture of itinerant electrons perceiving fluctuating confinement potentials, $T_2^*$ turns out to be about 20 ns, which is about 2 times longer than the hyperfine-limited dephasing times of shallow donors [22] and more in line with magneto-optical data for conduction band electrons in bulk Ge [53]. It shall be noted that in the latter case the spin decoherence time was found to be anisotropic, reflecting the intervalley scattering regime [39]. Fig.4b demonstrates that when the intravalley relaxation is the dominating process the ensemble dephasing time is not $g$-factor dependent and thus isotropic.

In the following we will try to extract the spin-lattice or longitudinal relaxation time $T_1$ from the power ($P$) dependence of continuous wave ESR [48]. To this end, we carried out selected measurements in a cylindrical cavity with high Q factor and offering a finite electric field of the microwave within the sample. Moreover, we restricted ourselves to the analysis of CESR lines at $g = g_t$, because, as shown before, those are unaffected by the inhomogeneous broadening induced by the interface roughness.

Fig.5a shows a colour-coded map of the CESR intensity as a function of $P$ in the $-30$ dB (low $P$) to $-7$ dB (high $P$) range for the resonance peak corresponding to degenerate C and D valleys measured at θ = 90° in the sample having 17 nm thick QWs. Besides readily demonstrating that $\Delta B_{pp}$ remains constant, Fig.5a shows that at low $P$ the CESR signal possesses the well-known absorption lineshape (AS), which results from spin-flip processes induced by the resonance between the microwave photons and the Zeeman splitting of the spin states. For a direct inspection, the CESR peak measured at $-30$dB is shown as a black line in the inset in Fig.5b. Surprisingly, however, we do not recover in our samples the typical increase and saturation with $P$ of the AS intensity that is routinely observed in electron spin resonance experiments [48]. Instead, a more puzzling behaviour can be appreciated in Fig.5a. At low $P$ the lineshape resembles the well-studied Dysonian shape observed in metals when dispersion of the microwave power arises because of skin effects at the metal surface [48]. The pattern is asymmetric because of the occurrence of an additional dispersion signal (DS), which in 2DEGs was reported for the first time in Si QWs and explained by considering the real component of the magnetic susceptibility of the samples [51]. Indeed, by increasing $P$ at first the AS becomes weak and at $P \sim -12$ dB, the lineshape gets fully modified, showing one unexpected negative dip, which stems from a pure DS (see also inset of Fig.5b). Notably, by further increasing $P$ the intensity of the resonance peak turns out to be strongly enhanced and the lineshape changes again showing this time a parity inversion with respect to the AS-like pattern of the low power regime (see also inset of Fig.5b). Such sign change of the absorption component compares well with the polarization signal (PS) occurring in 2D conduction electrons because of variations of the spin-dependent conductivity during the microwave absorption process[51].

In light of this discussion, the overall behaviour of the CESR lineshape as a function of $P$ can be accounted for by a linear superposition of the three AS, DS, and PS contributions (see Supplementary Section E). According to the model put forward in Ref. 51, the latter leads to a peak-to-peak amplitude $A_{pp}$ that scales as $\sim\sqrt{P^3}$, while AS and DS are both proportional to $\sqrt{P}$. Fig.5b, where we assumed negative amplitudes for PS-like peaks, shows that such phenomenological power law well describes our findings as AS (PS) dominates at low (high) $P$, while AS and PS cancel each other in the intermediate regime, eventually making the DS component clearly visible at $P \sim -12$dB.

As detailed in Supplementary Section E, modelling the resonance lines by these three signal components provides us with the $T_1$ and $T_2^*$ times summarised in Fig.5c for all the QW samples. For the sample with the thickest QWs, the model gives $T_2^*$ in good agreement with those anticipated in Fig.4b for all the $g$ factors, further corroborating our previous linewidth analysis. Fig.5c also shows that $T_2^*$ decreases and its values at 90° and 35° get closer in thinner QWs. This behaviour compares well with an enhancement in the electron localization when the QW width is reduced, and with the correspondingly increasing efficiency in the spin dephasing due to hyperfine coupling. Above all, Fig.5c discloses $T_1$ up to 5μs, thus more than two orders of magnitude longer than $T_2^*$. It is worth noting that such values are exceedingly longer than the one reported for conduction electrons in bulk Ge (see Ref. 23 and refs. therein). Such lengthening of the spin lifetime stems from the intravalley nature of the spin-flip processes addressed in our experiments, as opposed to the intervalley scattering regime probed in previous literature work. In addition, these findings point towards a mitigated role of impurity scattering [53].

Finally, $T_1$ of 2DEGs confined in Ge QWs strikingly matches spin relaxation times measured in Si QWs at comparable lattice temperatures [54]. Such result sheds light on the extraordinary coexistence of long spin relaxation times and large SOI in Ge, which unveils itself as an excellent candidate for the exploitation of spin currents in novel transport architectures, such as spin-based interconnects [55], transistors [56], and reprogrammable logic [57].

Our work based on modulation doped heterostructures provided us with the first electron spin resonance measurements that exquisitely address conduction electrons in Ge, thus granting direct access to intrinsic spin-dependent phenomena. CESR data of Ge QWs revealed unmatched tailoring of the electron Landé $g$ factor in group IV materials, and offered experimental insights into the entanglement between SOI and valley degrees of freedom. These unique spin properties of Ge can enrich group IV spintronics and enable new applications in quantum technologies. Moreover, our findings point out that the 2DEG can be surprisingly accompanied by a $g$ tensor mimicking the one of bulk material, a result that might stimulate further experimental and theoretical investigations. Looking ahead, 2DEGs in Ge can offer a special framework for quantum computation, opening unexplored pathways for future studies of spin physics in gate-defined low dimensional structures on Si.

**Acknowledgements**

We acknowledge T. Maggi and S. Bietti for assistance with the AFM measurements, L. Golub, M. M. Glazov and M. Guzzi for fruitful discussions. This work was supported by the Fondazione Cariplo through Grant SearchIV No. 2013-0623.

## References

1. Zutic, I., Fabian, J. & Das Sarma, S., Spintronics: Fundamentals and applications. *Rev. Mod. Phys.* **76**, 323-410 (2004).

2. Bychkov, Y. A. & Rashba, E. I., Properties of a 2D electron gas with lifted spectral degeneracy. *Z. Eksp. Teor. Fiz. Pis'ma* **39**, 78-81 (1984).

3. Manchon, A., Koo, H. C., Nitta, J., Frolov, S. M. & Duine, R. A., New perspectives for Rashba spin–orbit coupling. *Nature Mater.* **14**, 871-882 (2015).

4. Bychkov, Y. A. & Rashba, E. I., Oscillatory effects and the magnetic susceptibility of carriers in inversion layers. *J. Phys. C: Solid State Phys.* **17**, 6039-6045 (1984).

5. Bernevig, B. A., Orenstein, J. & Zhang, S.-C., Exact SU(2) Symmetry and Persistent Spin Helix in a Spin-Orbit Coupled System. *Phys. Rev. Lett.* **97**, 236601 (2006).

6. Koralek, J. D. *et al.*, Emergence of the persistent spin helix in semiconductor quantum wells. *Nature* **458**, 610-613 (2009).

7. Walser, M. P., Reichl, C., Wegscheider, W. & Salis, G., Direct mapping of the formation of a persistent spin helix. *Nature Phys.* **8**, 757-762 (2012).

8. D'yakonov, M. I. & Perel, V. I. *Pis'ma Z. Eksp. Teor. Fiz.* **13**, 657 (1971).

9. D'yakonov, M. I. & Perel, V. I., Current-induced spin orientation of electrons in semiconductors. *Phys. Lett. A* **35**, 459-460 (1971).

10. Sinova, J., Valenzuela, S. O., Wunderlich, J., Back, C. H. & Jungwirth, T., Spin Hall effects. *Rev. Mod. Phys.* **87**, 1213-1259 (2015).

11. Breton, J.-C. L., Sharma, S., Saito, H., Yuasa, S. & Jansen, R., Thermal spin current from a ferromagnet to silicon by Seebeck spin tunnelling. *Nature* **475**, 82–85 (2011).

12. Bauer, G. E. W., Saitoh, E. & Wees, B. J. v., Spin caloritronics. *Nature Mater.* **11**, 391–399 (2012).

13. Flatté, M. E. & Awschalom, D. D., Challenges for semiconductor spintronics. *Nature Phys.* **3**, 153 - 159 (2007).

14. Jansen, R., Silicon spintronics. *Nature Mater.* **11**, 400–408 (2012).

15. Wilson, D. K. & Feher, G., Electron Spin Resonance Experiments on Donors in Silicon. III. Investigation of Excited States by the Application of Uniaxial Stress and Their Importance in Relaxation Processes. *Phys. Rev.* **124**, 1068-1083 (1961).

16. Zutic, I., Fabian, J. & Erwin, S. C., Spin Injection and Detection in Silicon. *Phys. Rev. Lett.* **97**, 026602 (2006).

17. Wilamowski, Z., Jantsch, W., Malissa, H. & Roessler, U., Evidence and evaluation of the Bychkov-

Rashba effect in SiGe/Si/SiGe quantum wells. *Phys. Rev. B* **66**, 195315 (2002).

18. Tyryshkin, A. M., Lyon, S. A., Jantsch, W. & Schäffler, F., Spin Manipulation of Free Two-Dimensional Electrons in Si/SiGe Quantum Wells. *Phys. Rev. Lett.* **94**, 126802 (2005).

19. Matsunami, J., Ooya, M. & Okamoto, T., Electrically Detected Electron Spin Resonance in a High-Mobility Silicon Quantum Well. *Phys. Rev. Lett.* **97**, 066602 (2006).

20. Wilamowski, Z., Hans Malissa, H., Schaeffler, F. & Jantsch, W., g-Factor Tuning and Manipulation of Spins by an Electric Current. *Phys. Rev. Lett.* **98**, 187203 (2007).

21. Roth, L. M. & Lax, B., g FACTOR OF ELECTRONS IN GERMANIUM. *Phys. Rev.* **3**, 217-219 (1959).

22. Sigillito, A. J. *et al.*, Electron spin coherence of shallow donors in natural and isotopically enriched. *Phys. Rev. Lett.* **115**, 247601 (2015).

23. Giorgioni, A., Vitiello, E., Grilli, E., Guzzi, M. & Pezzoli, F., Valley-dependent spin polarization and long-lived electron spins in germanium. *Appl. Phys. Lett.* **105**, 152404 (2014).

24. Pillarisetty, R., Academic and industry research progress in germanium nanodevices. *Nature* **479**, 324–328 (2011).

25. Moriya, R. *et al.*, Cubic Rashba Spin-Orbit Interaction of a Two-Dimensional Hole Gas in a Strained-Ge/SiGe Quantum Well. *Phys. Rev. Lett.* **113**, 086601 (2014).

26. Ares, N. *et al.*, Nature of Tunable Hole g Factors in Quantum Dots. *Phys. Rev. Lett.* **110**, 046602 (2013).

27. Katsaros, G. *et al.*, Hybrid superconductor–semiconductor devices made from self-assembled SiGe nanocrystals on silicon. *Nature Nanotech.* **5**, 458-464 (2010).

28. Hu, Y., Kuemmeth, F., Lieber, C. M. & Marcus, C. M., Hole spin relaxation in Ge–Si core–shell nanowire qubits. *Nature Nanotech.* **7**, 47-50 (2012).

29. Maier, F., Klinovaja, J. & Loss, D., Majorana fermions in Ge/Si hole nanowires. *Phys. Rev. B* **90**, 195421 (2014).

30. Rieger, M. M. & Vogl, P., Electronic-band parameters in strained Si(1-x)Ge(x) alloys on Si(1-y)Ge(y) substrates. *Phys. Rev. B* **48**, 14276 (1993).

31. Virgilio, M. & Grosso, G., Type-I alignment and direct fundamental gap in SiGe based heterostructures. *J. Phys.: Condens. Matter* **18**, 1021-1031 (2006).

32. Kuo, Y.-H. *et al.*, Strong quantum-confined Stark effect in germanium quantum-well structures on silicon. *Nature* **437**, 1334-1336 (2005).

33. Chaisakul, P. *et al.*, Integrated germanium optical interconnects on silicon substrates. *Nature Photon.* **8**, 482–488 (2014).

34. Baron, F. A. *et al.*, Manipulating the L-valley electron g factor in Si-Ge heterostructures. *Phys. Rev. B*

**68**, 195306 (2003).

35. Pontinen, R. E. & Sanders, T. M. J., NEW ELECTRON SPIN RESONANCE SPECTRUM IN ANTIMONY-DOPED GERMANIUM. *Phys. Rev. Lett.* **5**, 311-313 (1960).

36. Hale, E. B., Dennis, J. R. & Pan, S.-H., Strain effects on the ESR spectrum from antimony donors in germanium. *Phys. Rev. B* **12**, 2553-2561 (1975).

37. Mitsuma, T. & Morigaki, K., Effects of Uniaxial [110]-Compressive Stress on the Additional Spin Resonance Spectrum in Sb-Doped Ge. *J. Phys. Soc. Jpn.* **20**, 491-499 (1965).

38. Roth, L. M., g Factor and Donor Spin-Lattice Relaxation for Electrons in Germanium and Silicon. *Phys. Rev.* **118**, 1534-1540 (1960).

39. Hautmann, C. & Betz, M., Magneto-optical analysis of the effective g tensor and electron spin decoherence in the multivalley. *Phys. Rev. B* **85**, 121203(R) (2012).

40. Li, P., Song, Y. & Dery, H., Intrinsic spin lifetime of conduction electrons in germanium. *Phys. Rev. B* **86**, 085202 (2012).

41. Wilson, D. K., Electron Spin Resonance Experiments on Shallow Donors in Germanium. *Phys. Rev.* **134**, A265-A286 (1964).

42. Tang, J.-M., Collins, B. T. & Flatté, M. E., Electron spin-phonon interaction symmetries and tunable spin relaxation in silicon and germanium. *Phys. Rev. B* **85**, 045202 (2012).

43. Li, P., Li, J., Qing, L., Dery, H. & Appelbaum, I., Anisotropy-Driven Spin Relaxation in Germanium. *Phys. Rev. Lett.* **111**, 257204 (2013).

44. Ivchenko, E. L. & Kiselev, A. A., Electron g factor of quantum wells and superlattices. *Sov. Phys. Semicond.* **26**, 827-831 (1992).

45. Salis, G., Kato, Y., Ensslin, K., Driscoll, D. C. & Gossard, A. C., Electrical control of spin coherence in semiconductor nanostructures. *Nature* **414**, 619-622 (2001).

46. Zinovieva, A. F. *et al.*, Electron localization in Ge/Si heterostructures with double quantum dots detected by an electron spin resonance method. *Phys. Rev. B* **88**, 235308 (2013).

47. Lipps, F. *et al.*, Electron spin resonance study of Si/SiGe quantum dots. *Phys. Rev. B* **81**, 125312 (2010).

48. Poole, C. P. J., *Electron Spin Resonance: a comprehensive treatise on experimental technique* (Dover Publications, 1996).

49. Feher, G., Wilson, D. K. & Gere, E. A., Electron spin resonance experiments on shallow donors in germanium. *Phys. Rev. Lett* **3**, 25-28 (1959).

50. Golub, L. E. & Ivchenko, E. L., Spin splitting in symmetrical SiGe quantum wells. *Phys. Rev. B* **69**, 115333 (2004).

51. Wilamowski, Z. & Jantsch, W., Suppression of spin relaxation of conduction electrons by cyclotron motion. *Phys. Rev. B* **69**, 035328 (2004).

52. Brower, K. L., Strain broadening of the dangling-bond resonance at the (111)Si-Si02 interface. *Phys. Rev. B* **33**, 4471-4478 (1986).

53. Lohrenz, J., Paschen, T. & Betz, M., Resonant spin amplification in intrinsic bulk germanium: Evidence for electron spin lifetimes exceeding 50 ns. *Phys. Rev. B* **89**, 121201 (2014).

54. Wilamowski, Z. *et al.*, Spin relaxation and g-factor of two-dimensional electrons in Si/SiGe quantum wells. *Physica E* **16**, 111-120 (2003).

55. Dery, H., Song, Y., Li, P. & Zutić, I., Silicon spin communication. *Appl. Phys. Lett.* **99**, 082502 (2011).

56. Datta, S. & Das, B., Electronic analog of the electrooptic modulator. *Appl. Phys. Lett.* **56**, 665-667 (1990).

57. Dery, H., Dalal, P., Cywinski, L. & Sham, L. J., Spin-based logic in semiconductors for reconfigurable large-scale circuits. *Nature* **447**, 573-576 (2007).

58. Birner, S. *et al.*, Nextnano: General purpose 3-D simulations. *TEEE Trans. Electron Devices* **54**, 2137–2142 (2007).

59. Paul, D. J., 8-band k.p modeling of the quantum confined Stark effect in Ge quantum wells on Si substrates. *Phys. Rev. B* **77**, 155323 (2008).

60. Teherani, J. T. *et al.*, Extraction of large valence-band energy offsets and comparison to theoretical values for strained-Si/strained-Ge type-II heterostructures on relaxed SiGe substrates. *Phys. Rev. B* **85**, 205308 (2012).

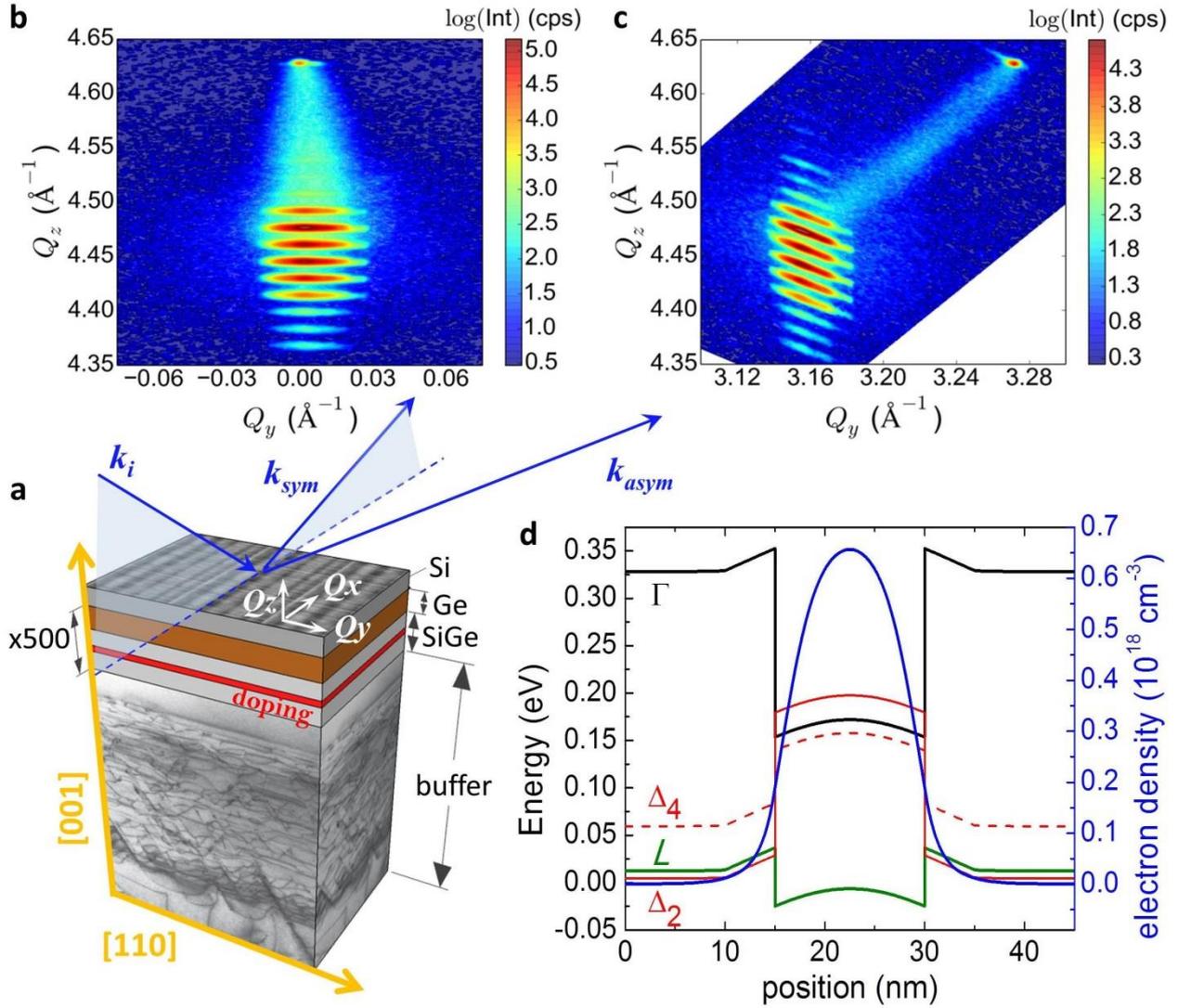

**Figure 1: a** Sketch of the structure (not to scale) of the n-type (P atoms) modulation doped Ge/SiGe QWs samples. Each sample consists of a 500-fold stack of QWs grown on (001)Si substrates. **b, c** Symmetric (004) and asymmetric (224) XRD reciprocal space maps of the sample with 20 nm QWs, respectively. The colour scale bar represents the XRD intensity. **d** Calculated conduction band alignment and electron density in the Ge/SiGe multiple QW structure.

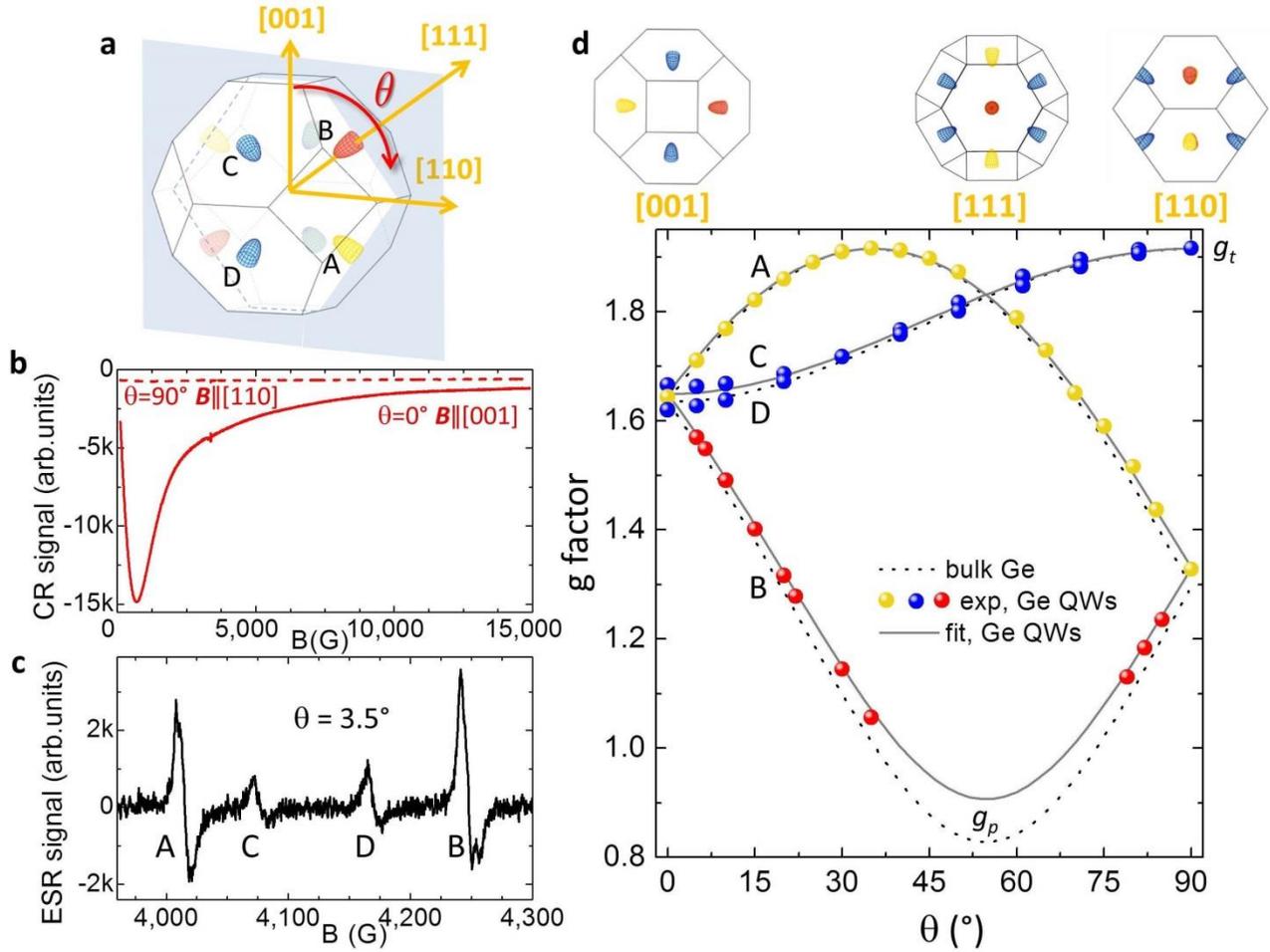

**Figure 2: a** Brillouin zone of bulk Ge. θ is the angle between the [001] crystallographic direction and the magnetic field **B**, which scans towards the [110] direction. The ellipsoidal isoenergetic surfaces of the conduction band at the *L* point are also shown. **b** Cyclotron Resonance (CR) signal in Ge QWs measured at T = 2 K for θ = 0° (solid red line) and θ = 90° (dashed red line). **c** ESR signals from conduction electrons in Ge QWs measured at T = 2 K and θ = 3.5°, after subtracting a linear background. **d** Values of *g* factor measured from the ESR peaks at T = 2 K in 20±1 nm Ge QWs as a function of θ. The correspondence between the angle θ and the main crystallographic directions are highlighted in the upper part of the figure. Labels from A to D establish the relation of the branches both with the peaks in panel **c** and with the valleys in panel **a**. $g_p$ and $g_t$ are the lowest and the highest values of *g*, respectively. Upper part: sketches of the Brillouin zone when **B** is directed along the main crystallographic axis.

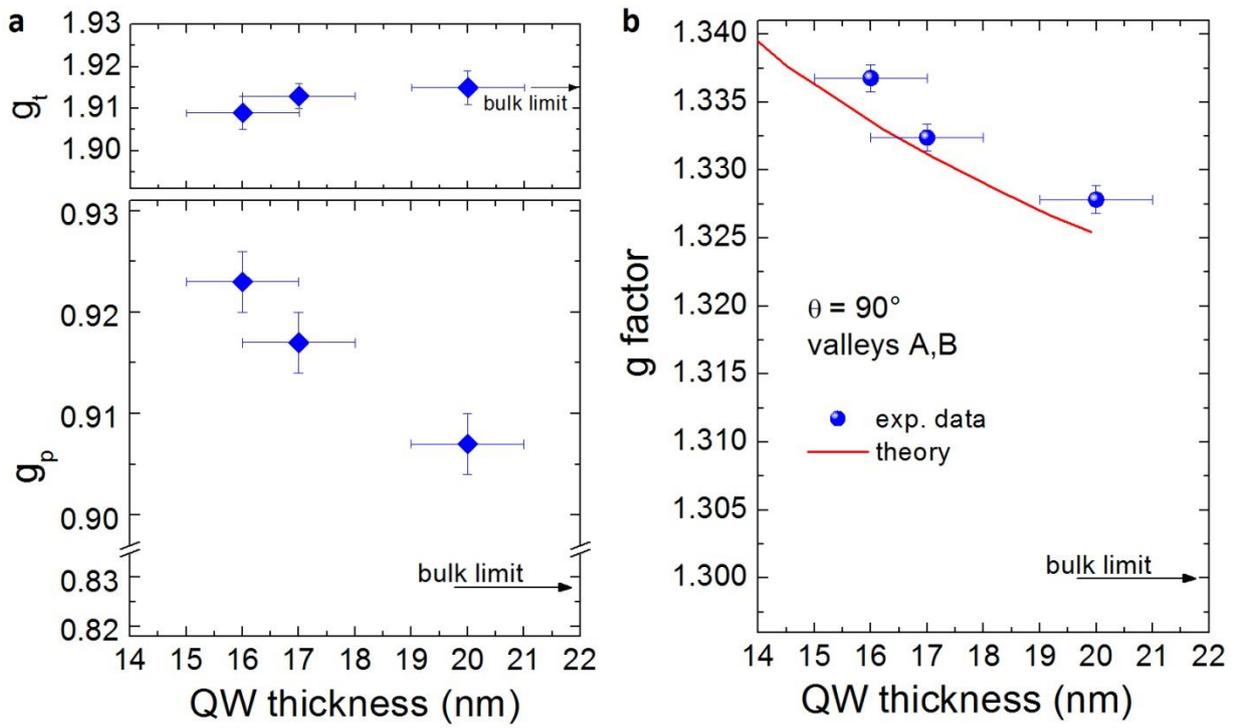

**Figure 3: a** Values of $g_p$ and $g_t$ parameters reported in Ref. 36 for bulk Ge (arrows), and obtained in this work (diamonds) for Ge QWs with different thicknesses. **b** *g* factor values at θ = 90° for the valleys A and B as a function of the QW thickness. The experimental data for Ge QWs (dots) are reported along with the values calculated by Baron et al. in Ref. 34 (red line), and the reference values of bulk Ge according to Ref. 36.

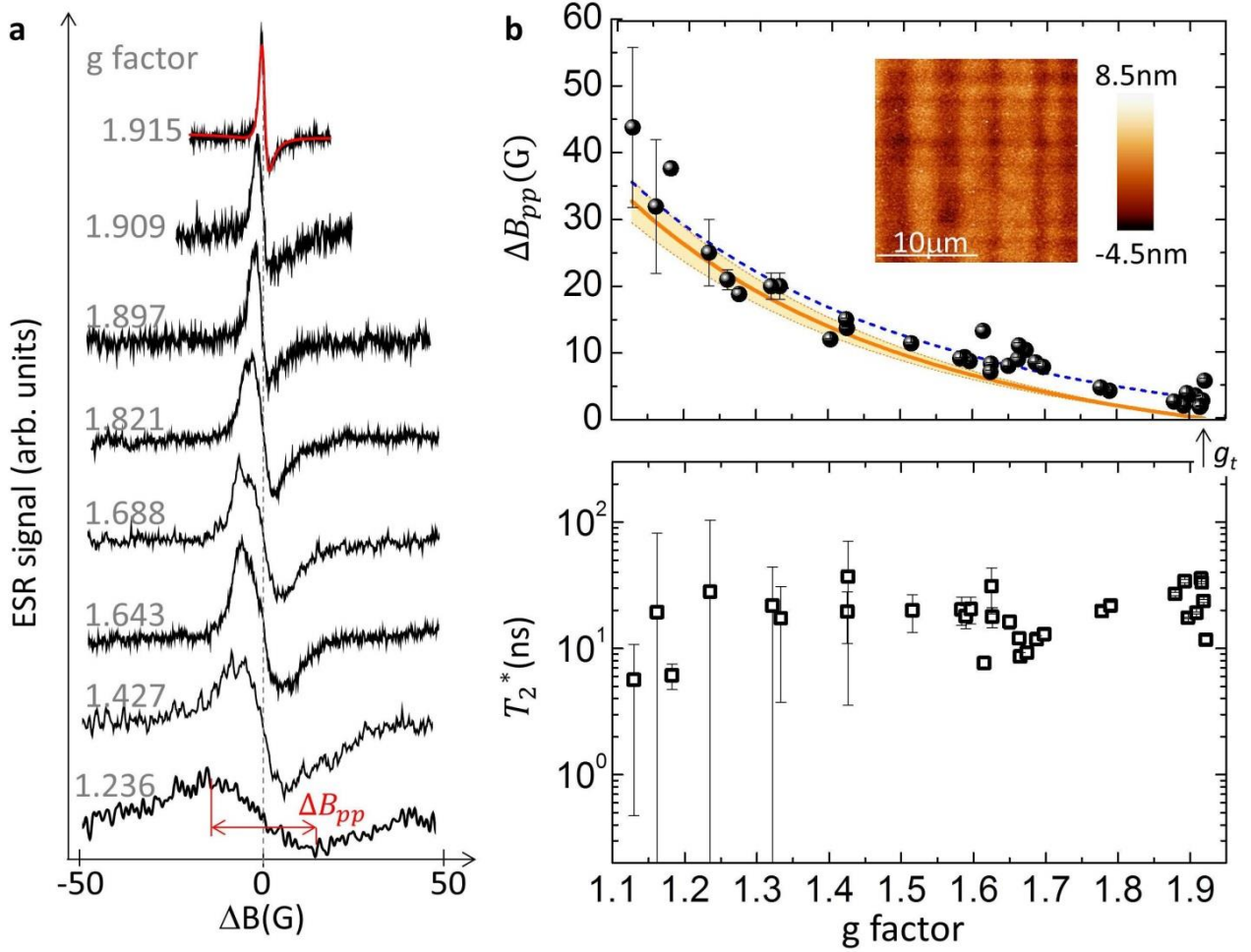

**Figure 4: a** CESR peaks for valleys A and B of the 20 nm Ge QWs observed at the different $g$ factors. The abscissa of each peak is shifted by an amount equal to the resonance field, and the spectra are vertically shifted for clarity. The peak-to-peak linewidth $\Delta B_{pp}$ of the ESR line at the smallest $g$ factor is shown. The ESR line at $g = 1.915$ is reported together with its Lorentzian fit (red line). **b** Upper panel: Full dots are the measured $\Delta B_{pp}$ for all the valleys of Fig.2 as a function of $g$, the orange line is the calculated inhomogeneous broadening $\Delta B^G{}_{pp}$ due to the interface roughness in 20 nm Ge QWs. Shadowed area corresponds to the error bar of the calculated $\Delta B^G{}_{pp}$, resulting from the roughness error. The dashed blue line corresponds to the orange curve, shifted by 2.9 G. Inset: AFM image of the surface of the sample having 20 nm thick QWs. Lower panel: $T_2^*$ values obtained by the linewidth of the ESR lines, after removing the inhomogeneous $\Delta B^G{}_{pp}$ broadening contribution due to the fluctuations of the QW width.

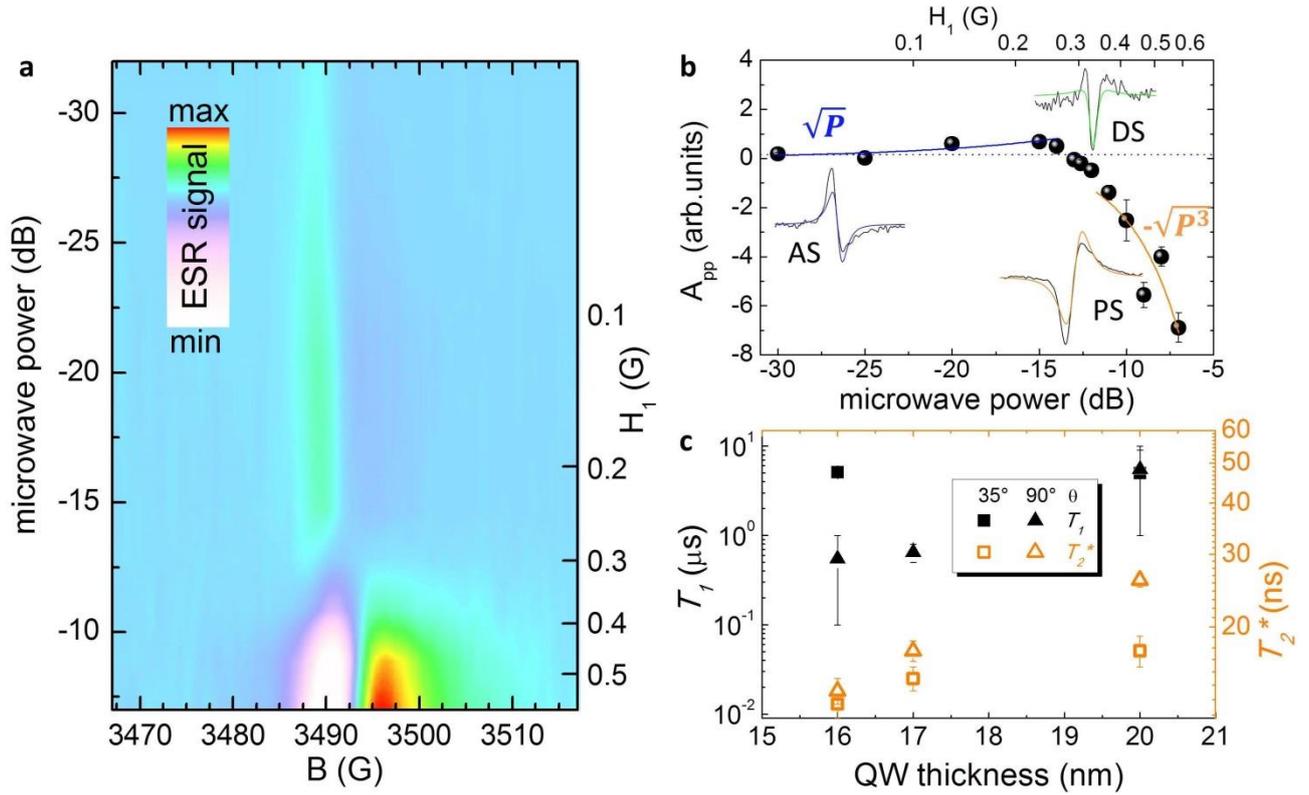

**Figure 5: a** Colour-coded intensity map of the CESR peak at $g = g_t$ and θ = 90° in 17 nm Ge QWs, as a function of the microwave power ($P$). $H_1$ is the intensity of the magnetic field of the microwave within the sample. **b** Peak-to-peak amplitude ($A_{pp}$) of the peaks shown in **a** vs $P$. The blue and orange solid lines are guides to the eye. The shape of the peak at -30 dB, -12.5 dB, and -7 dB are reported (black lines) together with the fitting curves (coloured lines) showing absorption (AS), dispersion (DS), and polarization (PS) components [51], respectively. $A_{pp}$ is considered positive when the ESR peak is AS-like, and negative when it is PS-like. **c** Values of spin-lattice relaxation time $T_1$ and ensemble dephasing time $T_2^*$ obtained from ESR peaks at $g = g_t$ in QWs with different thicknesses.